\documentclass[twocolumn,aps,prb,showpacs,amsmath,superscriptaddress]{revtex4}
\usepackage{amssymb}
\usepackage{mathrsfs}
\usepackage{graphicx}
\usepackage{float}
\usepackage{amsmath}

\begin{document}
\title{Dynamical signature of localization-delocalization transition in one-dimensional incommensurate lattice}
\author{Chao Yang}
\affiliation{Beijing National Laboratory for Condensed
Matter Physics, Institute of Physics, Chinese Academy of Sciences, Beijing 100190,
China}
\affiliation{School of Physical Sciences, University of Chinese Academy of Sciences, Beijing, 100049, China}

\author{Yucheng Wang}
\affiliation{Beijing National Laboratory for Condensed
Matter Physics, Institute of Physics, Chinese Academy of Sciences, Beijing 100190,
China}
\affiliation{School of Physical Sciences, University of Chinese Academy of Sciences, Beijing, 100049, China}

\author{Pei Wang}
\affiliation{Department of Physics, Zhejiang Normal University, Jinhua 321004, China}

\author{Gao Xianlong}
\affiliation{Department of Physics, Zhejiang Normal University, Jinhua 321004, China}

\author{Shu Chen}
\thanks{Corresponding author: schen@iphy.ac.cn}
\affiliation{Beijing National Laboratory for Condensed
Matter Physics, Institute of Physics, Chinese Academy of Sciences, Beijing 100190,
China}
\affiliation{School of Physical Sciences, University of Chinese Academy of Sciences, Beijing, 100049, China}
\affiliation{Collaborative Innovation Center of Quantum Matter, Beijing, China}
\date{ \today}

\begin{abstract}
We investigate the quench dynamics of a one-dimensional incommensurate lattice described by the Aubry-Andr\'{e} model by a sudden change of the strength of incommensurate potential $\Delta$ and unveil that the dynamical signature of localization-delocalization transition can be characterized by the occurrence of zero points in the Loschmit echo. For the quench process with quenching taking place between two limits of $\Delta=0$ and $\Delta=\infty$, we give analytical expressions of the Loschmidt echo, which indicate the existence of a series of zero points in the Loschmidt echo. For a general quench process, we calculate the Loschmidt echo numerically and analyze its statistical behavior. Our results show that if both the initial and post-quench Hamiltonian are in extended phase or localized phase, Loschmidt echo will always be greater than a positive number; however if they locate in different phases, Loschmidt echo can reach nearby zero at some time intervals.
\end{abstract}


\maketitle
\section{Introduction}
In recent years, dynamical quantum phase transition (DQPT) has extended our understanding of phase transitions and universality greatly \cite{Heyl,CK,FA,MM,Jpg,HeylD,HeylS,MSc,HeylT,Eck,Aaz,ZH}, which provides us a new perspective on exploring the behavior of phase transitions far from equilibrium. As a simple but important paradigm of nonequilibrium processes, the quantum quench attracted intensive studies. To describe the dynamics of a quantum system which is pushed out equilibrium by a sudden change of the Hamiltonian, an important quantity is the Loschmidt echo, which measures the overlap of the initial quantum state and the time-evolved state after the quench \cite{LE,SunCP,LE2,Jafari}. Many theoretical works have demonstrated that the Loschidmt echo plays an important role in characterizing the nonequilibrium dynamic signature of a quantum phase transition \cite{Heyl,SunCP,CK,Eck}. After mapping the Loschmidt amplitude to a boundary partition function, the singularity of dynamical free energy density in thermodynamic limit can be found at critical times $\{t^{*}\}$, which are similar to the well-known Fisher zeros \cite{Fish}. This singularity is found to have relationship with the dynamics of order parameter \cite{HeylD}. Up to now, the DQPT has been explored in a series of models which are known to exhibit quantum phase transition, such as transverse field Ising model (TFIM) \cite{Heyl}, anisotropic XY model \cite{DoraXY,Jmh}, Hubbard and  Falicov-Kimball models \cite{Eck} and two-band topological systems \cite{HeylT,Dora,SSh,UB}, etc. Thanks to the developments of quantum simulation techniques, DQPT has been realized by ions simulations of TFIM \cite{Cfr}. Besides, by observing the appearance, movement, and annihilation of vortices in reciprocal space, dynamical topological order parameter has also been recognized in optical lattices systems \cite{Cw}.

According to the theory of DQPT, the appearance of a series of zero points in the Loschmidt echo at critical times $\{t^{*}\}$ can be viewed as a dynamic signature of quantum phase transitions. While most theoretical studies of DQPT and Loschmidt echo focus on the traditional quantum systems driven by competing quantum terms \cite{Heyl,CK,FA,MM,Jpg,HeylD,HeylS,MSc,HeylT,Eck,Aaz}, less attention is paid on the dynamics and Loschmidt echo in a quantum disorder system which exhibits the localization-delocalization transition \cite{Pwa}. A natural but interesting question is whether the Loschmit echo can still be used  to characterize the DQPT of a quantum disorder system? And if yes, whether we can observe zero points of the Loschmidt echo by studying the quench dynamics of the quantum disorder system? Aiming to answer these questions, in this work we shall study the quench dynamics in a one-dimensional (1D) incommensurate lattice, which is effectively described by the Aubry-Andr\'{e} (AA) model \cite{AA,Modugno}. It is known that all the eigenstates of the AA model are either extended or localized and there exists a transition from an extended state to a localized state with the change of the strength of incommensurate potential \cite{AA,Modugno,Machida,Geisel,Roscilde,Roux,Albert,Cai,Gullo}. The localization transition in the 1D incommensurate lattice has been experimentally verified in a bichromatic optical lattice by observing the expansion dynamics of a Bose-Einstein condensate initially trapped in the center of optical lattice \cite{Roati}, which exhibits different diffusion behaviors for the extended or localized phase \cite{Tg,Tgs,Lsch,Tk,Qin}. Different from previous works on the expansion dynamics, we study the quench dynamics with the initial state being an eigenstate of the initial Hamiltonian. After performing a sudden quench of the strength of incommensurate potential $\Delta$, the behaviors of Loschmidt echo are found to be quite different depending on whether the initial and final Hamiltonians locate in the same phase regime or not.


The paper is organized as follows. In section II, we introduce the model and study the quench dynamics in the limiting cases of quenching between $\Delta=0$ ($\Delta \rightarrow\infty$) and $\Delta \rightarrow\infty$ ($\Delta=0$). In these two limiting cases, we can give analytical expressions of the Loschmidt echo and demonstrate that there are a series of zeros of Loschmidt echo, which is also consistent with our numerical results. In section III, we study the general quench process, for which no analytical results are available and we thus study the evolution of Loschmidt echo with the help of numerical methods. By analyzing the statistical behavior of the values of Loschmidt echo in a long time, we demonstrate that Loschmidt echo will oscillate and take a finite value in the case of quenching in the same phase, while Loschmidt echo can approach zero in the case of quenching in different phases. A brief summary is given in section IV.

\section{Model and quench dynamics}
We investigate the AA model with Hamiltonian
\begin{equation}\label{Eq1}
H(\Delta)=-J\sum_{i=1}^{N}(c_i^\dagger c_{i+1}+h.c.)+\Delta\sum_{i=1}^{N} \cos(2\pi\alpha i)c_i^{\dagger}c_i,
\end{equation}
where $c_i^{\dagger}(c_i)$ denotes the creation (annihilation) operator of fermions, $J$ is hopping amplitude, $\alpha$ is an irrational number and $\Delta$ is the strength of the incommensurate potential. The incommensurate potential can be viewed as a kind of quasi-random disorder, which drives the system undergoing a delocalization-localization transition at $\Delta=2J$. When $\Delta<2J$, all the eigenstates are extended, whereas all the eigenstates are localized, when $\Delta>2J$ \cite{AA}. For convenience we take $J=1$ as the energy unit and fix $\alpha=\frac{\sqrt{5}-1}{2}$.

While conventional studies of dynamical properties in disordered systems focus on the diffusion of a wave packet, in this work we consider the quench dynamics of the incommensurate system described by the AA model. By preparing the system as an eigenstate of the Hamiltonian $H(\Delta_i)$ and then performing a sudden quench to the final Hamiltonian $H(\Delta_f)$, we consider the behavior of Loschmidt amplitude (return amplitude)
\begin{align}\label{Eq2}
G(t,\Delta_i,\Delta_f) = \langle\Phi_0(\Delta_i)|e^{-itH(\Delta_f)}|\Phi_0(\Delta_i)\rangle,
\end{align}
and Loschmidt echo (return probability)
\begin{align}\label{Eq3}
L(t,\Delta_i,\Delta_f)=|G(t,\Delta_i,\Delta_f)|^2,
\end{align}
where $|\Phi_0(\Delta_i)\rangle$ stands for the eigenstate of the initial Hamiltonian, and $\Delta_i$ ($\Delta_f$) represents the strength of the incommensurate potential corresponding to the initial (final) state before (after) the quench.

It is known that the Loschmidt echo plays an important role in the theory of DQPTs.  The behavior of Loschmidt echo approaching zero at some times $t$ in the thermodynamic limit can be viewed as a signature of the occurrence of the DQPT, which has been demonstrated by studying various models exhibiting quantum phase transitions. However, it is still not clear whether the Loschmidt echo approaching zero can be viewed as a signature for the localization-delocalization transition, which shall be clarified in this work. In order to get an intuitive understanding, we first consider two limiting cases of quench processes, i.e., quench processes between states with $\Delta_i=0$ $(\infty)$ and $\Delta_f=\infty$ $(0)$, which can be calculated analytically, whereas the general quench processes between arbitrary $\Delta_i$ and $\Delta_f$ are studied with the help of numerical calculations..
\begin{figure}[htbp]
    \centering
    \includegraphics[width=\linewidth]{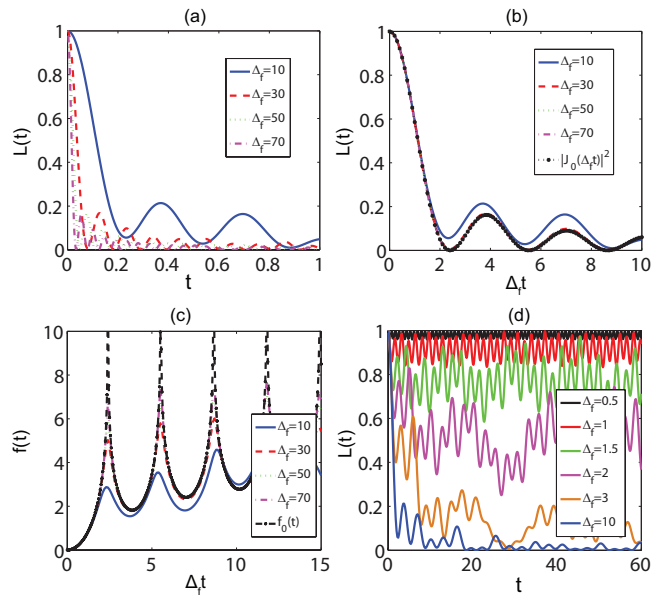}\\
    \caption{(Color online) Evolution of Loschmidt echo with different $\Delta_f$s and the size of the system $N=1000$. The initial state is fixed to be the ground state of the initial Hamiltonian with $\Delta_i=0$. (a) $L$ versus $t$. (b) $L$ versus the rescaled time $\Delta_f t$. (c) Evolution of ``dynamic free energy" $f(t)$ for various large $\Delta_f$s. The black dashed-dotted curve corresponds to the analytical result $f_0(t)=-\log |J_0(\Delta_f t)|^2$. (d) Evolution of Loschmidt echo for various $\Delta_f$s. For $\Delta_f>2$, $L(t)$ will approach zero after some time intervals.} \label{Fig1}
\end{figure}

In the first case, we fix $\Delta_i=0$ and consider the periodic boundary condition, i.e., the system is initially prepared in a plane wave state, which is the eigenstate of the Hamiltonian (\ref{Eq1}) with $\Delta_i=0$:
\begin{equation}
|\phi_k\rangle= \frac{1}{\sqrt{N}} \sum_{j=1}^{N} e^{i k j}c_j^{\dagger}|0\rangle,  \label{Eq5}
\end{equation}
where the wave vector $k=\frac{2\pi (l-N/2)}{aN}\in (-\frac{\pi}{a},\frac{\pi}{a}]$ ($l=1,\cdots,N$) lies in the first Brillouin zone (BZ) and $a$ represents the lattice spacing. The corresponding eigenvalue is
\begin{equation}
E(k)=2\cos ka \label{Eq4} .
\end{equation}
Performing a sudden quench of $\Delta$ from the initial value  $\Delta_i$ to the final value $\Delta_f$, the return amplitude can be written as
\begin{align}\label{Eq6}
  \nonumber G_k(t) &= \langle\phi_k|e^{-iH(\Delta_f)t}|\phi_k\rangle =  \sum_m \langle\phi_k|e^{-iH(\Delta_f)t}|m\rangle\langle m|\phi_k\rangle\\
    & = \sum_m e^{-iE_m t}|\langle m|\phi_k\rangle|^2,
\end{align}
where $E_m$ and $|m\rangle$ denote the $m-th$ eigenevalue and eigenstate of final Hamiltonian respectively.

Now use the fact that when in the limit of $\Delta_f \rightarrow \infty$, the energy is determined by the diagonal terms and the eigenstates are localized in a single site, we can simplify Eq. (\ref{Eq6}) and obtain
\begin{align}
  \nonumber G_k(t) = \frac{1}{N}\sum_{m=1}^{N} e^{-i\Delta_f t\cos(2\pi\alpha m)} .
\end{align}
For an irrational number $\alpha$, the phase $2\pi\alpha m$  ($m=1,\cdots,N$) modulus $2\pi$ is distributed randomly between $-\pi$ and $\pi$ when we sum over $m$ to the large $N$ limit. So we can approximately replace the summation by the integration
\begin{align}\label{Eq7}
  G_k(t) \approx \frac{1}{2\pi}\int_{-\pi}^{\pi}e^{-i\Delta_f t\cos\theta}d\theta = J_0(\Delta_f t),
\end{align}
where $J_0(\Delta_f t)$ is the zero-order Bessel function. From the properties of Bessel function, we know that $J_0(x)$ has a series of zeros $x_\alpha$ with $\alpha=1, 2, 3, \cdots$. These zeros indicate the Loschmit amplitude and echo reach zero at times
\begin{equation}\label{Eq8}
t_\alpha^{*}=\frac{x_\alpha}{\Delta_f}.
\end{equation}

According to the theory of DQPT, the occurrence of a series of zeros in the Loschmit amplitude can be viewed as the signature of DQPT as these zeros correspond to divergences of the boundary partition function. Although the analytical result is obtained in the limit of $\Delta_f \rightarrow \infty$, the above results are expected to hold true as long as $\Delta_f$ is large enough (see Fig. \ref{Fig1}(a)-(c)). Since the transition time $t_\alpha^{*}$ is inversely proportional to $\Delta_f$, the Loschmidt echo will oscillate more rapidly as $\Delta_f$ is increasing. If we rescale the time $t\rightarrow\Delta_f t$, the evolution of Loschmidt echo will display a similar behavior for quench processes with different $\Delta_f$. To see it clearly, we display the numerical results of the evolution of Loschmidt echo as a function of $t$ and $\Delta_f t$ in Fig. \ref{Fig1}(a) and Fig. \ref{Fig1}(b), respectively. Here the initial strength is set at $\Delta_i=0$. It is clear that the Loschmit echoes $L(t)$ for $\Delta_f=30$, $50$ and $70$ oscillate with different frequencies, but they almost completely overlap to the analytical result $|J_0(\Delta_f t)|^2$ and are indistinguishable after rescaling the time as shown in Fig. \ref{Fig1}(b). When $\Delta_f=10$, the Loschmit echo obviously deviates $|J_0(\Delta_f t)|^2$, indicating the analytical result obtained in the limit $\Delta_f \rightarrow \infty$  is no longer a good approximation. To see the zeros of $L(t)$ more clearly, we can use the ``dynamical free energy" which is defined as $f(t)=-\log L(t)$ \cite{Heyl}, where $f(t)$ will be divergent at the dynamical phase transition time $t=t_\alpha^{*}$. The evolution of $f(t)$ for different $\Delta_f$s are shown in Fig. \ref{Fig1}(c). $L(t)$ exhibits obvious peaks around $t=t_\alpha^{*}$ and the behavior gets more close to the limiting case with the increasing of $\Delta_f$.

In Fig. \ref{Fig1}(d), we display $L(t)$ versus $\Delta_f t$ for various $\Delta_f$ with $\Delta_f=10$, $3$, $2$, $1.5$, $1$ and $0.5$ from bottom to top.
It can be seen that $L(t)$ exhibits different behaviors for  $\Delta_f>2$ and  $\Delta_f<2$. For $\Delta_f<2$, $L(t)$ oscillates with its average decreasing with the increasing of $\Delta_f$. We do not find any zero point of $L(t)$ even in a long time, which is obviously different from cases with $\Delta_f>2$. As a comparison, for the case of $\Delta_f=3$ the Loschmit echo $L(t)$ has an obvious decay and reaches nearby zero at about $\Delta_f t = 28.5$. With the increase of $\Delta_f$, $L(t)$ decays more quickly and gets more closed to the limiting case described by Eq. (\ref{Eq7}).
\begin{figure}[htbp]
    \centering
    \includegraphics[width=\linewidth]{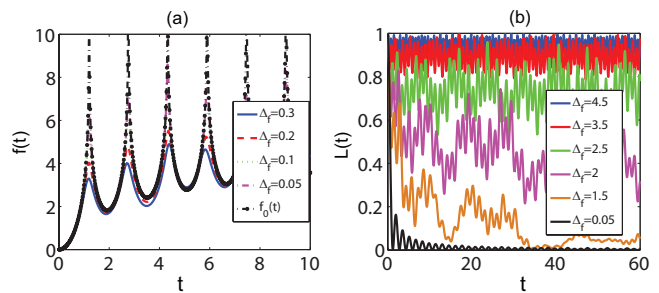}\\
    \caption{(Color online) (a) Evolution of ``dynamical free energy" $f(t)$ for large $\Delta_f$s. The black dashed-dotted curve corresponds to the analytical result $f_0(t)=-\log |J_0(2t)|^2$. (b) Evolution of Loschmidt echo with various $\Delta_f$s and the size of the system $N=1000$. The initial state is fixed to be the ground state of the initial Hamiltonian with $\Delta_i=100$. }\label{Fig2}
\end{figure}

Next we consider the quench processes from a very large $\Delta_i$ to $\Delta_f=0$. In the limit of $\Delta_i \rightarrow \infty$, the initial state is chosen as an eigenstate of the system, which is localized in one site, e.g., the site $m$. Substituting Eq. (\ref{Eq4}) and (\ref{Eq5}) into Eq. (\ref{Eq2}) we get
\begin{align}
  \nonumber G_m(t) &= \langle m|e^{-iH(\Delta_f)t}|m\rangle  = \sum_k \langle m|e^{-iH(\Delta_f)t}|\phi_k\rangle\langle\phi_k|m\rangle \\
  \nonumber  & = \sum_k e^{-2it\cos ka}|\langle m|\phi_k\rangle|^2  = \frac{1}{N}\sum_k e^{-2it\cos ka} .
\end{align}
In the large N limit, we can replace the summation by the integration, which leads to
\begin{align} \label{Eq9}
  \nonumber G_m(t) & = \frac{a}{2\pi}\int_{-\frac{\pi}{a}}^{\frac{\pi}{a}}e^{-2it\cos ka}dk \\
    & = J_0(2t).
\end{align}
From this expression, it is clear that the zeros of Loschmit echo occur at
\begin{equation}\label{Eq10}
t_\alpha^{*}=\frac{x_\alpha}{2},
\end{equation}
which are half of the zeros of the zero-order Bessel function $J_0(x)$. When $\Delta_f$ deviates a little from the limit case of $\Delta_f=0$, the analytical result Eq.(\ref{Eq9}) is expected to be still a good approximation.
Different from Eq. (\ref{Eq8}), the transition time $t_\alpha^{*}$ is independent of $\Delta_f$. Furthermore, we find that $t_\alpha^{*}$ is also not sensitive to the initial value $\Delta_i$ as long as $\Delta_i$ is large enough because the only information of the initial Hamiltonian we have used is the localized wave function.

In Fig. \ref{Fig2}, we show the numerical results for quenching processes with the initial state prepared in a localized state, which is taken to be the ground state of the initial system with $\Delta_i=1000$. From Fig. \ref{Fig2} (a), we see that systems with $\Delta_f=0.3$, $0.2$, $0.1$ and $0.05$ display similar behaviors to the limit case with $\Delta_f=0$, for which the divergent points of $f(t)$ occur at $t=t_\alpha^{*}$.  The more close to $\Delta_f=0$, the curves of numerical results are more close to the analytical result $f_0(t)=-\log |J_0(2t)|^2$, which are not sensitive to the values of $\Delta_f$. In Fig. \ref{Fig2}(b), we display $L(t)$ versus $t$ for various $\Delta_f$ with $\Delta_f=0.05$, $1.5$, $2$, $2.5$, $3.5$ and $4.5$ from bottom to top.
Similar to the previous case displayed in Fig. \ref{Fig1}(d), $L(t)$ exhibits quite different behavior for  $\Delta_f>2$ and  $\Delta_f<2$. For $\Delta_f<2$, $L(t)$ will approach zero at some given times. On the other hand, when $\Delta_f>2$, $L(t)$ never approaches zero in the evolution process.

\section{Numerical study of a general quench process}
\begin{figure}[htbp]
    \centering
    \includegraphics[width=\linewidth]{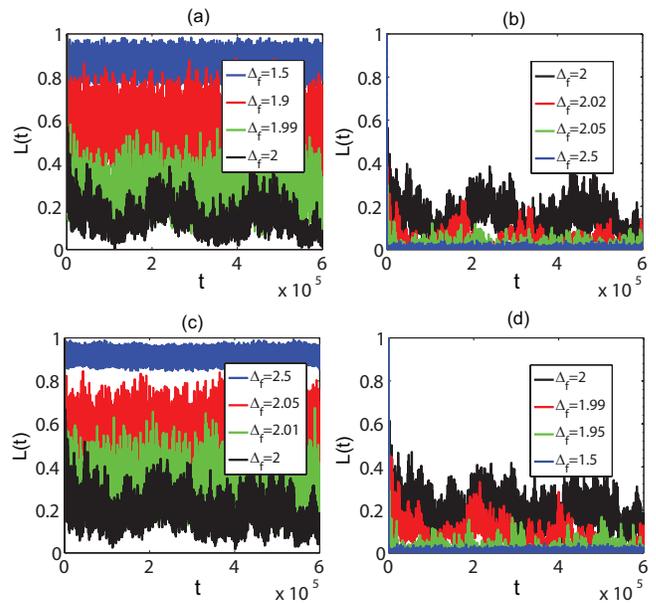}\\
    \caption{(Color online) Evolution of Loschmidt echo in a long time $T=6\times10^5$. The initial state is chosen to be the ground state of the initial Hamiltonian with (a) (b) $\Delta_i=0.5$ and (c) (d) $\Delta_i=4$. Different $\Delta_f$s are shown by different colors. Loschmidt echo can reach nearby zero only if $\Delta_f$ passes through the critical point $\Delta=2$.}\label{Fig3}
\end{figure}

In the above section, starting from the initial state prepared in the limit case with $\Delta_i=0$ (or $\Delta_i \rightarrow \infty$), we have shown that the Loschmidt echo can reach nearby zero in the evolution process if the incommensurate strength $\Delta_f$ after the quench is larger (or less) than the critical value $\Delta_c=2$, which is also the localization-to-delocalization transition point of the AA model. Now we consider the general cases that $\Delta_i$ and $\Delta_f$ are neither close to zero nor the infinity limit. Although no analytical solution can be found for the general case, we can still explore whether the presence or absence of the zeros of Loschmidt echo can still serve as a characteristic signature of dynamic quantum phase by numerically analyzing the evolution of the Loschmidt echo. In Fig. \ref{Fig3}, we show the evolution of Loschmidt echo for various $\Delta_f$s with $\Delta_i=0.5$ in (a) and (b), and $\Delta_i=4$ in (c) and (d), respectively. If both $\Delta_i$ and $\Delta_f$ locate in the same regime, i.e., both in the regime of $\Delta>2$ or  $\Delta<2$, $L(t)$ oscillates and has a positive lower bound, which never approaches zero during the evolution process, as shown in  Fig. \ref{Fig3} (a) and (c). However, if $\Delta_i$ and $\Delta_f$ locate in different regimes, $L(t)$ shall approach zero after some time intervals, as shown in  Fig. \ref{Fig3} (b) and (d).

\begin{figure}[htbp]
    \centering
    \includegraphics[width=0.9\linewidth]{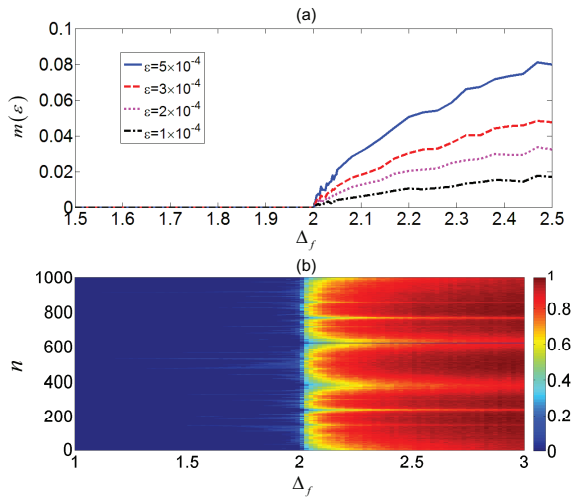}\\
    \caption{(Color online) The behavior of $m$ as a function of $\Delta_f$ for $N=1000$, $T=6\times10^{5}$ and $\Delta_i=0.5$. (a) Different $\epsilon$s are shown by different colors. There is a sharp increasing around $\Delta_f=2$. It is shown that $m=0$ for $\Delta_f<2$ and $m>0$ for $\Delta_f>2$. (b) Different choice of initial state with $n$ standing for the label of eigenstates of $H(\Delta_i)$. A clear boundary located at $\Delta_f=2$ can be seen. Here $\epsilon=0.01$.}\label{Fig4}
\end{figure}

To give a quantitative description on how Loschmidt echo approaches zero, we define a cutoff of small value $\epsilon$ close to zero. At a given large length range of time $T$, we measure the length of time interval which fulfills $L(t)\leq \epsilon$ in $t \in [0,T]$. Denoting this length as $M(\epsilon)$, which is a function of $\epsilon$ when fixing $T$, it can be viewed as the Lebesgue measure $I(L\leq \epsilon)$ \cite{Sim}. For convenience we use a normalized function $m(\epsilon)=\frac{M(\epsilon)}{T}$.
In Fig. \ref{Fig4}(a) we show $m (\epsilon)$ as a function of $\Delta_f$ for different $\epsilon$s with $\Delta_i=0.5$ fixed in the extended regime. Here the initial state is chosen as the ground state of the system. It can be seen that the behavior of $m(\epsilon)$ is quite different for $\Delta_f<2$ and $\Delta_f>2$. For $\Delta_f<2$, $m(\epsilon)$ is always zero for $\epsilon=5 \times 10^{-4}$, $3 \times 10^{-4}$, $2\times 10^{-4}$ and $1\times 10^{-4}$. However, there is a sharp increasing as $\Delta_f$ passes through the transition point $\Delta_c=2$, and $m (\epsilon)$ takes a finite value when $\Delta_f>2$.  Despite the fact that the value of $m (\epsilon)$ in the regime of $\Delta_f>2$ depends on the cutoff value $\epsilon$, we note that the sharp change behaviors around the transition point are similar for different cutoffs.

Although the initial state is taken to be the ground state of $H(\Delta_i)$ in the above calculations, we would like to indicate that our conclusion is independent of the choice of the initial eigenstates. To see this clearly,  in Fig. \ref{Fig4}(b) we show $m(\epsilon)$ as a function of $\Delta_f$ by choosing different eigenstates of $H(\Delta_i)$ as the initial state with  $\Delta_i=0.5$ and $\epsilon=0.01$. We can see that there exists a clear boundary at $\Delta_f=2$. For $\Delta_f<2$, $m$ is close to zero in the whole region. A sharp increase can be found around the transition point $\Delta_c=2$ for all the initial eigenstates, and $m (\epsilon)$ takes a finite value when $\Delta_f>2$.
\begin{figure}[htbp]
    \centering
    \includegraphics[width=\linewidth]{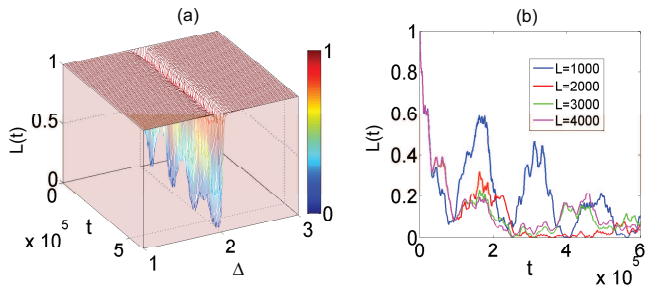}\\
    \caption{(Color online) (a) Loschmidt echo as a function of $\Delta$ and $t$ for the system with $N=1000$. The valley only occurs at the critical point $\Delta=2$. (b) The cross section of $\Delta=2$ for different sizes of systems.} \label{Fig5}
\end{figure}

Finally, we consider the special case of $\Delta_f$ being very close to $\Delta_i$. In such a case, the Loschmidt echo can be represented as
\begin{align}\label{Eq13}
L(t,\Delta,\delta) = |\langle\Phi_0(\Delta-\delta)|e^{-itH(\Delta+\delta)}|\Phi_0(\Delta-\delta)\rangle|^2,
\end{align}
where $\delta$ is a very small value. In terms of the above definition, a sharp decay of the Loschmidt echo around the critical point has been taken as the signature of quantum phase transition \cite{SunCP,Yuan,Zhong,Hamma,Sharma,Zanardi,Chen}. The quench process can be viewed as from $\Delta-\delta$ to $\Delta+\delta$, so the initial and final Hamiltonian are quite similar except around the critical point $\Delta=2$. In Fig. \ref{Fig5}(a), we fix $\delta=0.02$ and show the Loschmidt echo as a function of $\Delta$ and $t$. A deep valley can be found at $\Delta=2$, as the localization-delocalization transition enhances the decay of Loschmidt echo. While in the region apart from the critical point, the Loschmidt echo oscillates near $L(t)\sim 1$ and does not decay in a long time. The cross section of Fig. \ref{Fig5}(a) at $\Delta=2$ is shown in Fig. \ref{Fig5}(b). As a comparison, we also provide results for systems with different sizes. It is clear that the Loschmidt echo decays in an oscillating way and can always reach nearby zero in quite a long time interval, which is consistent with our conclusions.

\section{Conclusion}
In summary, we have studied the quench dynamics of the AA model by preparing the initial state as an eigenstate of the initial Hamiltonian  $H(\Delta_i)$ and  then performing a sudden quench to the final Hamiltonian $H(\Delta_f)$. For the quench process between two limiting cases, i.e., with $\Delta_i=0$ and $\Delta_f= \infty$ or $\Delta_i=\infty$ and $\Delta_f= 0$, we obtain the analytical expression of the Loschmidt echo, which suggests the existence of a series of zero points at critical times $\{t^{*}\}$. By comparing with the numerical results, we find the analytical results are still good approximations as long as the quench parameters deviate these limits not far away. For the general quench processes, we study the statistical behavior of Loschmidt echo numerically and demonstrate that Loschmidt echo would oscillate but never decay to zero in a long time if $\Delta_i$ and $\Delta_f$ are located in the same phase; however, Loschmidt echo would decay and reach nearby zero if $\Delta_i$ and $\Delta_f$ are located in different phases. Our results suggest that the occurrence of zero points in the Loschmidt echo can give a dynamical signature of localization-delocalization transition in the 1D incommensurate lattice.

\begin{acknowledgments}
The work is supported by the National Key Research and Development Program of China (2016YFA0300600), NSFC under Grants No. 11425419, No. 11374354 and No. 11174360, and the Strategic Priority Research Program (B) of the Chinese Academy of Sciences  (No. XDB07020000).
\end{acknowledgments}

\end{document}